\documentclass[aps,onecolumn,notitlepage,noshowpacs,noshowkeys]{revtex4-1}
\usepackage{amssymb, amsmath}
\usepackage{color}
\usepackage{graphicx}

\begin{document}
\title{Tunable particle separation via deterministic absolute negative mobility}
\author{A. S{\l}apik}
\affiliation{Institute of Physics, University of Silesia, 40-007 Katowice, Poland}
\author{J. Spiechowicz}
\email[Correspondence should be addressed to J.S.]{ (e-mail: jakub.spiechowicz@us.edu.pl)}
\affiliation{Institute of Physics, University of Silesia, 40-007 Katowice, Poland}
%
\begin{abstract}
Particle isolation techniques are in the spotlight of many areas of science and engineering. In food industry, a harmful bacterial activity can be prevented with the help of separation schemes. In health care, isolation techniques are used to distinguish cancer and healthy cells or in therapy for Alzheimer's and Parkinson's diseases. We consider a cloud of Brownian particles of different sizes moving in a periodic potential and subjected to an unbiased driving as well as a constant force. We reveal an efficient separation strategy via the counterintuitive effect of negative mobility when particles of a given size are transported in a direction opposite to the applied constant force. We demonstrate a tunable separation solution in which size of the particle undergoing separation may be controlled by variation of the parameters of the external force applied to the system. This approach is an important step towards the development of point-of-care lab-on-a-chip devices.
\end{abstract}
\maketitle
\section{Introduction}
Separation of (sub)micro sized particles is of paramount importance due to its vast applications including in particular medical diagnostics \cite{yager2006}. Anomalies in a bioparticle size often indicate various illnesses. This is apparent, for instance, for Alzheimer's, Parkinson's \cite{korecka2007} and Huntington's disease \cite{heffner1978}, needless to say that often cancer cells noticeably deviates in size from healthy ones \cite{suresh2007}. A reliable and effective approach to separating bioparticles is therefore much in demand. They span a large size range covering several orders of magnitude from nanometers to micrometers \cite{bhagat2010,xuan2014} what crucially complicates the development of such highly anticipated strategies. For instance, deadly viruses like HIV or COVID-19 are approximately 0.1 micrometer in diameter \cite{fehr2015}. On the other hand, the soma of a neuron can vary from 4 to 100 micrometers in diameter \cite{kandel2000}. For such a broad (sub)micrometer scale, efficient isolation techniques are required to allow tunability of the particle size intended for separation. Unluckily, the latter are rather scarce \cite{sajeesh2014, sonker2019}, however, recently some progress has been made in this direction \cite{bogunovic2012,kim2018,zhang2018}

In this work we aspire to partially fill this significant know-how gap by demonstrating a nonintuitive, yet efficient separation strategy taking advantage of a paradoxical mechanism of negative mobility \cite{eichhorn2002, machura2007, spiechowicz2014pre,spiechowicz2019njp}. We show that under the action of a static bias only particles of a given linear size move in the direction opposite to this net force whereas the others migrate concurrently towards it. This effect creates a possibility of steering different species of particles in opposite directions under identical experimental conditions thus facilitating their separation. A proof-of-principle experiment of a similar isolation scheme was performed with insulator dielectrophoresis in a nonlinear, symmetric microfluidic structure with electrokinetically activated transport \cite{ros2005, eichhorn2010}. Recently, such a setup allowed to induce the negative mobility not only for a colloidal particle but also for a biological compound in the form of mouse-liver mitochondria \cite{luo2016}.

Motivated by the large size range encountered in biochemical applications, a crucial result of this work is a demonstration of a tunable separation strategy, in which the size of the particle undergoing separation may be effectively controlled by variation of the parameters characterizing the external force applied to the particle, e.g. the magnitude of the static bias. The same setup can be applied to segregate particles with respect to their mass in a similar tunable manner \cite{slapik2019prl}. This approach may provide selectivity required for individual isolation of nano and micro particles, proteins, organelles and cells and thus constitutes an important step towards the development of robust lab-on-a-chip devices exploited in both research and industrial applications, in particular point-of-care medical diagnostics.

The paper is organized as follows. In Sec. II we outline the model of a Brownian particle dwelling in a spatially periodic potential under the action of both an external harmonic driving as well as a constant bias. In the next section we exemplify the negative mobility phenomenon. Sec. IV provides crucial results of the paper, namely, a tunable particle separation strategy. In Sec. V we discuss the possibility of tailoring particle isolation in the considered system. Finally, the last section is devoted to summary and conclusion.
\section{Model}
The system considered in this study is a classical inertial Brownian particle of mass $M$ which moves in a spatially periodic one-dimensional potential \mbox{$U(x) = U(x + L)$} of the period $L$, additionally subjected to an unbiased time-periodic force $A\cos{(\Omega t)}$ of the amplitude $A$ and the angular frequency $\Omega$, as well as an external static force $F$. Dynamics of such a particle is described by the Langevin equation \cite{hanggi2009}
\begin{equation}
\label{model}
	M\ddot{x} + \Gamma\dot{x} = -U'(x) + A\cos{(\Omega t)} + F + \sqrt{2\Gamma k_B T}\,\xi(t),
\end{equation}
where the dot and the prime denote differentiation with respect to time $t$ and the particle coordinate $x$, respectively. Coupling of the particle with thermal bath of temperature $T$ is modeled by Gaussian white noise of zero mean and unity intensity, namely
\begin{equation}
	\langle \xi(t) \rangle = 0, \quad \langle \xi(t) \, \xi(s) \rangle = \delta(t-s).
\end{equation}
The noise intensity factor $2\Gamma k_B T$ (where $k_B$ is the Boltzmann constant) follows from the fluctuation-dissipation theorem \cite{marconi2008fluctuation} and ensures that the system reaches the equilibrium state when \mbox{$A = 0$} and $F = 0$. The potential $U(x)$ is spatially periodic with the period $L$ and the barrier height $2\Delta U$,
\begin{equation}
	U(x) = \Delta U \sin \left(\frac{2\pi}{L}x\right).
\end{equation}
At first glance the studied system looks simply, however, it exhibits peculiar transport behaviour including noise-enhanced transport efficiency \cite{spiechowicz2014pre}, anomalous diffusion \cite{spiechowicz2017scirep,spiechowicz2019chaos}, amplification of normal diffusion \cite{reimann2001,spiechowicz2015chaos} and a non-monotonic temperature dependence of normal diffusion \cite{spiechowicz2020pre}.

As the first step of the analysis we transform the Eq. (\ref{model}) into its dimensionless form. This aim can be achieved in several ways. It often allows to simplify the setup description as after the rescaling procedure some parameters appearing in the corresponding dimensional version may be eliminated thus reducing the complexity of the problem. Moreover, recasting into the dimensionless variables ensures that the results obtained later are independent of the setup which is essential to facilitate the choice in realizing the best setup for testing theoretical predictions in experiments. Here, we propose the use of the following scales as the characteristic units of length and time
\begin{equation}
	\label{scaling}
	\hat{x} = \frac{x}{L}, \quad \hat{t} = \frac{t}{\tau_0}, \quad \tau_0 = L \sqrt{\frac{M}{\Delta U}}.
\end{equation}
The Langevin equation (\ref{model}) transformed according to the above variables reads
\begin{equation}\label{dimlessmodel}
    \ddot{\hat{x}} + \gamma \dot{\hat{x}} = - \hat{U}'(\hat{x}) + a \cos{(\omega \hat{t})} + f + \sqrt{2 \gamma D} \, \zeta(\hat{t}).
\end{equation}
The rescaled \textit{dimensionless friction coefficient} $\gamma$ is the ratio of two characteristic time scales
\begin{equation}
\label{gamma}
\gamma = \frac{\tau_0}{\tau_1} = \frac{\Gamma L}{\sqrt{M \Delta U}},
\end{equation}
where $\tau_1 = M/\Gamma$ is characteristic time for the velocity relaxation of the free Brownian particle. The dimensionless mass is set to unity, $m = 1$. Other rescaled parameters are as follows: \mbox{$a = (L/\Delta U) A$}, $\omega = \tau_0 \Omega$, \mbox{$f = (L/\Delta U)F$}. The dimensionless potential \mbox{$\hat{U}(\hat{x}) = U(L\hat{x})/\Delta U = \sin(2\pi \hat x)$} has the period $\hat L = 1$. 
The dimensionless thermal noise $\zeta(\hat{t})$ assumes the same statistical properties as $\xi(t)$, i.e. $\langle \zeta(\hat{t}) \rangle = 0$ and $\langle \zeta(\hat{t})\zeta(\hat{s}) \rangle = \delta(\hat{t} - \hat{s})$. The dimensionless noise intensity $D = k_BT/\Delta U$ is the ratio of thermal energy and half of the non-rescaled potential barrier. From now on, only the dimensionless variables will be used in this study and therefore, in order to simplify the notation, the $\wedge$-symbol will be omitted in the equation (\ref{dimlessmodel}).

The dimensionless friction coefficient $\gamma$ is the most important parameter for later mentioned process of the particle separation with respect to its size. It is due to the fact that even for the simplest model of hydrodynamic interactions occurring in this setup it depends on the linear size $R$ of the particle. For instance, the spherical particle diffusing in the surrounding medium is subjected to Stokes drag $-\Gamma \dot{x}$ where $\Gamma = 6 \pi \eta R$ and $\eta$ is the viscosity of the environment \cite{landau_hydro}. We stress that (sub)micro sized particles typically possess small mass and therefore for them the dimensionless friction coefficient $\gamma$ given by Eq. (\ref{gamma}) is expected to be either of the order or larger than the dimensionless mass $m = 1$.

\subsection{The directed velocity}
The particle mobility describes its ability to move through the medium as a response to the biased force acting on it. Hence, the observable of foremost interest in this study is a directed velocity $\langle v \rangle$ of the particle which may be written as 
\begin{equation}
\label{directedvelocity}
\langle v \rangle = \lim_{t \to \infty} \frac{1}{t} \int_0^{t} ds \, \langle \dot{x}(s) \rangle,
\end{equation}
where $\langle \cdot \rangle$ indicates averaging over all realizations of thermal noise as well as over initial conditions for the particle position $x(0)$ and its velocity $\dot{x}(0)$. The latter is obligatory for the deterministic limit $D \propto T \to 0$  when the dynamics may be non-ergodic and results can be affected by the specific choice of initial conditions \cite{spiechowicz2016scirep}.

The Fokker-Planck equation corresponding to the Langevin Eq. (\ref{dimlessmodel}) cannot be solved analytically in a closed form. Therefore the system may be analyzed only by numerical simulations. Equation (\ref{dimlessmodel}) is characterized by a 5-dimensional parameter space $\{\gamma, a, \omega, f, D\}$ the detailed exploration of which is a very challenging task. All numerical calculations were performed using an innovative computational method which is based on employing graphics processing unit supercomputers. This procedure allowed us to speed up computations by about $10^3$ times compared to traditional methods. For technical details we refer the reader to Ref. \cite{spiechowicz2015cpc}.
\begin{figure}[t]
	\centering
	\includegraphics[width=0.48\linewidth]{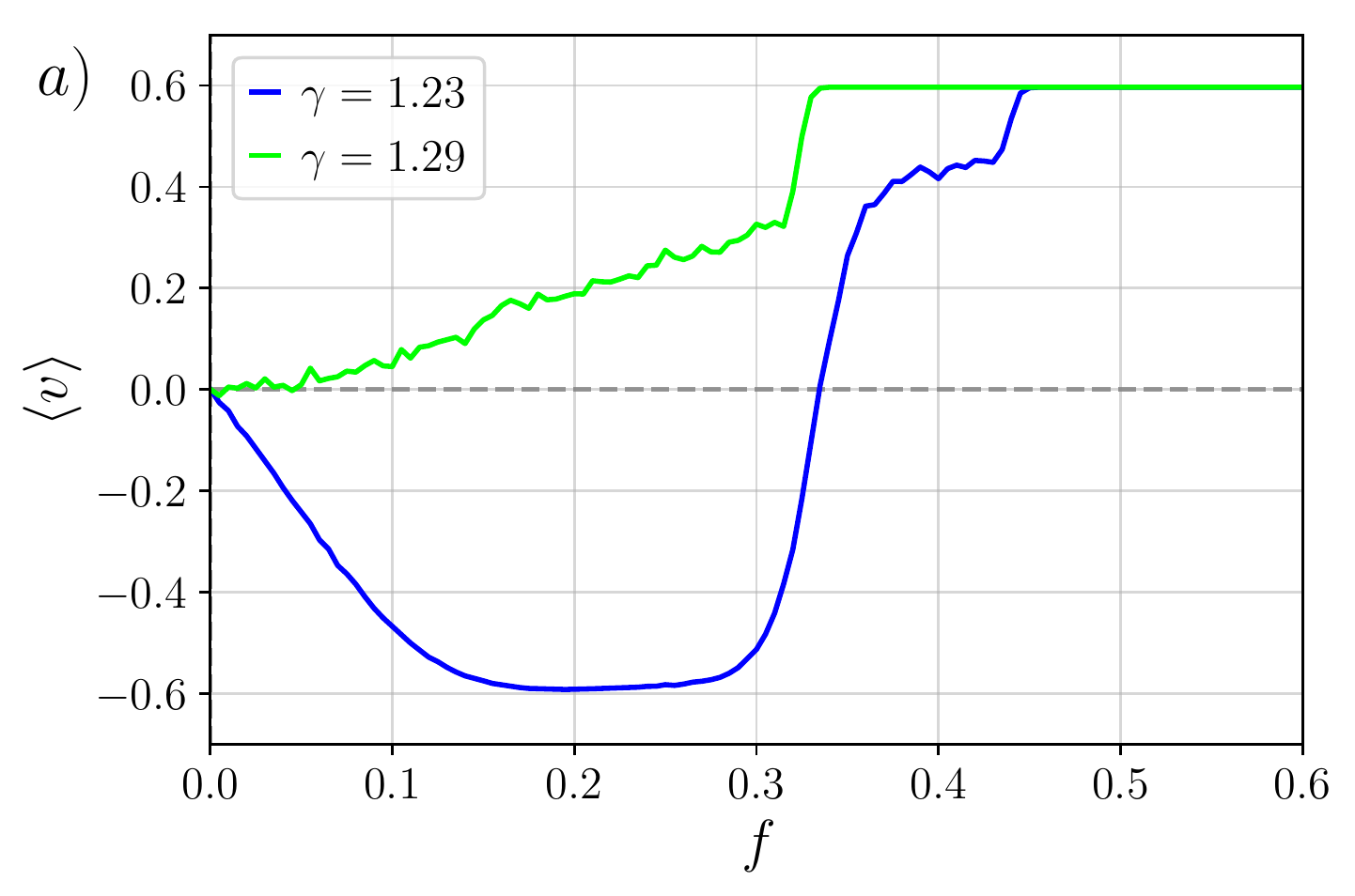}
	\includegraphics[width=0.49\linewidth]{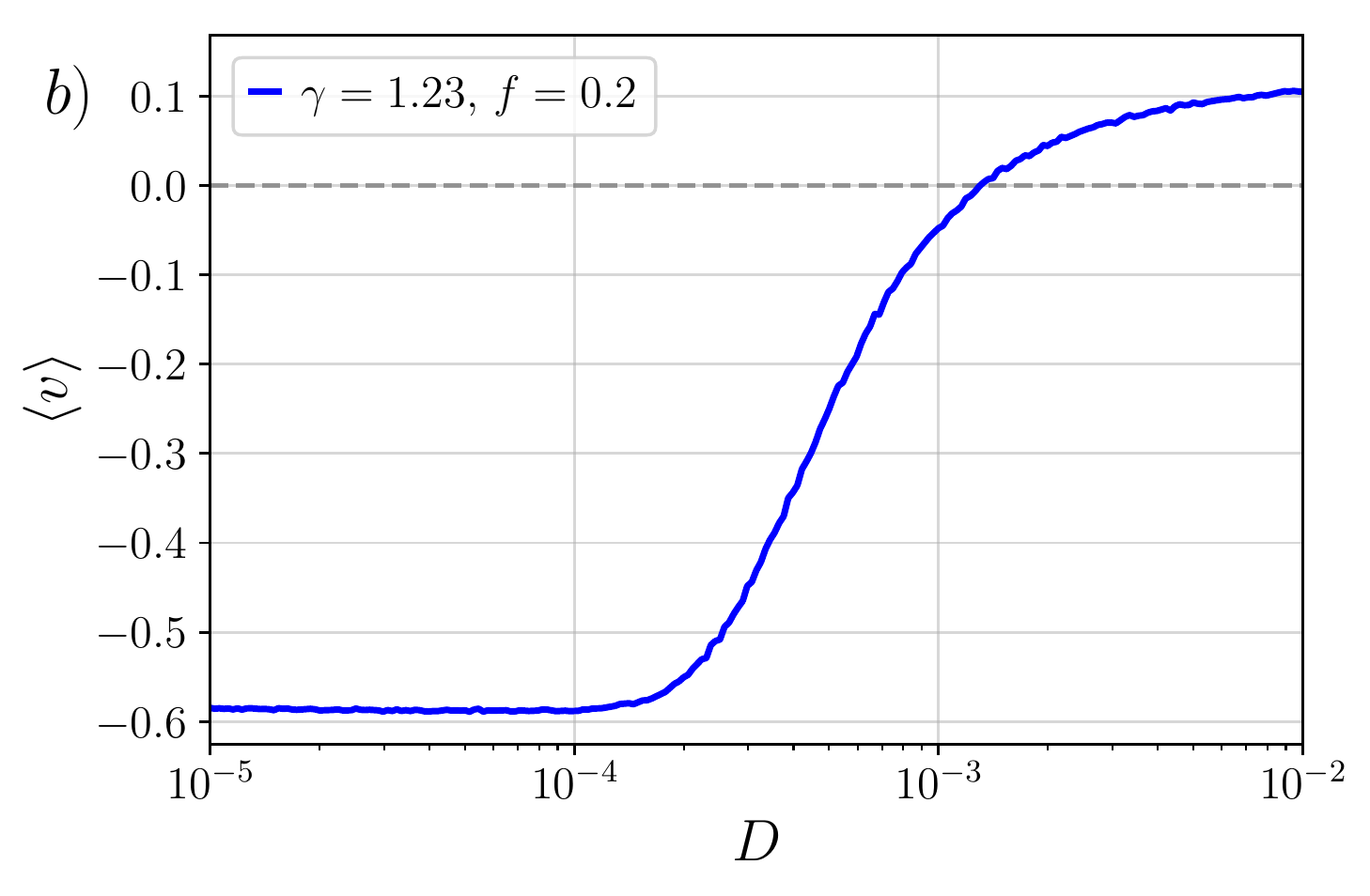}
	\caption{The average velocity $\langle v \rangle$ depicted as a function of the static force $f$ for two different values of the friction $\gamma$. Normal and anomalous transport behaviour in the form of negative mobility is observed for $\gamma = 1.29$ and $\gamma = 1.23$, respectively. In the panel (b) we present the average velocity $\langle v \rangle$ versus temperature $D \propto T$ for $\gamma = 1.23$. Other parameters are: $a = 4.5, \omega = 3.75$, $f = 0.2$, and $D = 0.0001$.} 
	\label{fig1}
\end{figure}
\section{Negative mobility}
In general, the directed velocity $\langle v \rangle$ is an increasing function of the static force $f$ and the resultant particle transport follows the direction of the bias, i.e. \mbox{$\langle v \rangle = \mu(f) f$} with a positive nonlinear mobility $\mu(f) > 0$. However, in the parameter space there are also regimes for which the particle moves on average in the opposite way, namely $\langle v \rangle < 0$ for $f > 0$. Such anomalous transport behaviour is characterized by a negative mobility $\mu(f) < 0$ \cite{machura2007,speer2007,nagel2008}. The key ingredient for the occurrence of the latter effect is that the system is driven away from thermal equilibrium into a time-dependent nonequilibrium state \cite{machura2007}. This fact provides a negation of the Le Chatelier-Braun equilibrium principle \cite{landau_stat} stating that the response of the perturbed system occurs into the direction of the applied bias towards a new equilibrium state. In our case this condition is guaranteed by the presence of the external harmonic driving $a\cos{(\omega t)}$.

It is known that there exist two fundamentally different mechanisms responsible for the emergence of negative mobility in the considered system. Firstly, it may be generated solely by the deterministic dynamics given by Eq. (\ref{dimlessmodel}) with $D \propto T = 0$ \cite{machura2007,speer2007}. Secondly, it can be induced by an appropriate dose of thermal fluctuations \cite{machura2007}. Among the first mentioned class we distinguish two completely distinct scenarios. For the deterministic counterpart of the system the negative mobility may be induced either by chaos-assisted dynamics \cite{speer2007} or regular, non-chaotic attractors transporting the particle in the direction opposite to the applied bias \cite{slapik2018cnsns}. Our numerical research reveals that the most common reason for the occurrence of the negative mobility is rooted in the complexity of the deterministic and chaotic dynamics \cite{slapik2018cnsns,slapik2019prappl}. This observation is of great importance for uncovering the parameter regimes allowing the particle separation.

In Fig. \ref{fig1} we illustrate the negative mobility effect. For $\gamma = 1.29$ the directed velocity $\langle v \rangle > 0$ assumes the same sign as the force $f > 0$ leading to the normal particle transport regime with $\mu > 0$. However, for $\gamma = 1.23$, the directed velocity $\langle v \rangle < 0$ is opposite to the bias $f > 0$ and in consequence the negative mobility effect emerges $\mu < 0$. If the value of $f$ is positive and large enough then the sign of $\langle v \rangle$ coincides with the force $f$ again. Therefore this anomalous transport behaviour is observed only in the vicinity of the zero bias $f = 0$ and therefore often is termed as the \emph{absolute} negative mobility \cite{kostur2008}. As it is illustrated in the the panel (b) the presented parameter regime belongs to the class of deterministically induced negative mobility as for the limiting case of vanishing thermal noise intensity $D \to 0$ the directed velocity is negative $\langle v \rangle < 0$. We want to stress that this limit should be considered with utmost care as in such a case attractors transporting the particle in opposite directions may coexist and the dynamics may be non-ergodic. It means that depending on the initial conditions, a particle would either move in the direction of the bias force or opposite to it \cite{speer2007}. Nevertheless, at any finite $D > 0$, possibly coexisting deterministic attractors turn metastable and due to thermally activated transitions between them the ergodicity of dynamics is restored. Consequently, the directed velocity $\langle v \rangle$ is independent of the initial conditions. Moreover, as it is shown in the panel (b) the deterministically induced negative mobility effect is generally quite robust with temperature change and usually survives up to moderate thermal noise intensities. 

\begin{figure}[t]
	\centering
	\includegraphics[width=0.49\linewidth]{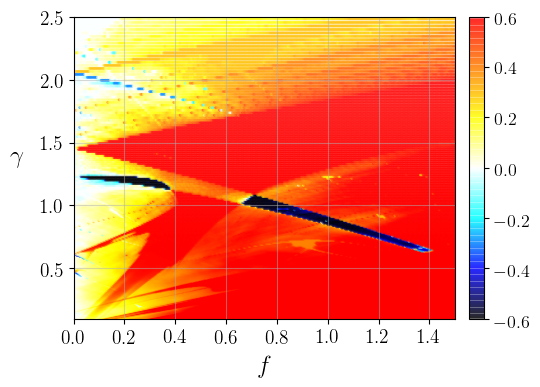}
	\caption{Two-dimensional map of the directed velocity $\langle v \rangle$ as a function of the static force $f$ and the friction coefficient $\gamma$. The magnitude of $\langle v \rangle$ is indicated by the color bar. Blue areas represent the parameter regimes corresponding to the negative mobility. Parameter values: $a=4.5$, $\omega = 3.75$ and $D = 0$.}
	\label{fig2}
\end{figure}
\section{Tunable particle separation}
We now want to harvest the negative mobility phenomenon to separate (sub)micrometer sized particles. This task may be achieved as follows. Imagine that there is a mixture of several species of spherical particles each differing by its linear size $R$. The friction coefficient $\gamma$, via e.g. Stokes formula, depends on the particle radius $\gamma = \gamma(R)$. We can extract the given species of particles $R^*$ characterized by the friction coefficient $\gamma^* \equiv \gamma(R^*)$ if only for this particular group of particles the negative mobility effect arises, i.e. $\langle v \rangle < 0$ for $\gamma^*$ and $\langle v \rangle > 0$ for the rest. Therefore this task translates to discovery in the complex four-dimensional parameter subspace $\{a, \omega, f, D\}$ regimes where in the characteristic $\langle v \rangle(\gamma)$ there exists only one interval $\delta \gamma$ of the friction coefficient around the desired particle size $\gamma^*$ for which the negative mobility $\langle v \rangle < 0$ emerges. In such a way only particles with the radius $R^*$ would be extracted from the mixture. 

Motivated by the large size range typically encountered in biochemical applications, we aim to develop a tunable scheme that allows to control the particle size targeted for isolation by changing only one parameter of the system. In \cite{slapik2019prl} it has been shown that the negative mobility effect can be harvested to separate the (sub)micrometer sized particles with respect to their mass. The particle mass targeted for isolation might be effectively controlled over a regime of nearly two orders of magnitude upon changing solely the frequency $\omega$ of the external harmonic driving. Moreover, in \cite{slapik2019prappl} an efficient separation mechanism based on thermal fluctuation induced negative mobility phenomenon has been proposed. By tuning solely temperature of the system $D \propto T$, one can extract from the mixture of particle species differing by size only those of a strictly defined radius. This scheme opens an opportunity to separate particles that carry no charge or dipole, however, it may be inconvenient since temperature variation typically takes too much time to offer a robust experimental implementation. Therefore, in contrast, in this work we harvest the negative mobility effect to develop the particle separation strategy in which the particle size intended for isolation will be controlled by changing only the parameters characterizing the externally applied force, namely, the static bias $f$ or the amplitude $a$ or the frequency $\omega$ of the harmonic driving.

Unfortunately, there is no clear relationship between the presence of the negative mobility and the model parameter values. A tiny displacement in the parameter space may either cause a sudden emergence of the negative mobility or its rapid decay. Therefore, extensive numerical simulations of Eq. (\ref{dimlessmodel}) were performed in order to systematically investigate the established parameter space. As the deterministically induced negative mobility is the most populated mechanism in the parameter space we set $D = 0$. Then Eq. (\ref{dimlessmodel}) was simulated for several values of the bias $f \in [0,2]$ and a parameter domain in the area $\gamma \times a \times \omega \in [0.1,10] \times [0,25] \times [0,20]$ at a resolution of 400 points per dimension. Overall, we considered nearly $10^9$ different parameter sets. This exceptional precision was possible only because of our innovative simulation method \cite{spiechowicz2015cpc}.

The so collected data was transformed into two-dimensional maps presenting the directed velocity $\langle v \rangle$ versus two chosen model parameters to facilitate the further analysis. The results of foremost interest are those with $\gamma$ dependence since the friction coefficient may be used as an indicator to differentiate particles by their size. We explored the data to discover any correlations between the presence of the negative mobility, the friction coefficient $\gamma$ as well as the magnitude of the parameters $a$, $\omega$ and $f$. We exemplify such a situation in Fig. \ref{fig2} where we depict the directed velocity $\langle v \rangle$ as a function of the static bias $f$ and the friction coefficient $\gamma$. The color bar in the plot represents the magnitude of the directed velocity $\langle v \rangle$. The occurrence of the negative mobility is marked by blue areas for which $\langle v \rangle < 0$. The reader can observe a linear trend between the friction coefficient $\gamma$, the static bias $f$ and the emergence of the anomalous transport. The negative mobility effect occurs for progressively smaller values of $\gamma$ as the force $f$ increases. Among many so discovered negative mobility regimes we distilled only those where the latter phenomenon is present solely for one indivisible interval $\delta \gamma$ of the friction coefficient $\gamma$ thus permitting the particle separation.
\begin{figure*}[t]
	\centering
	\includegraphics[width=0.49\linewidth]{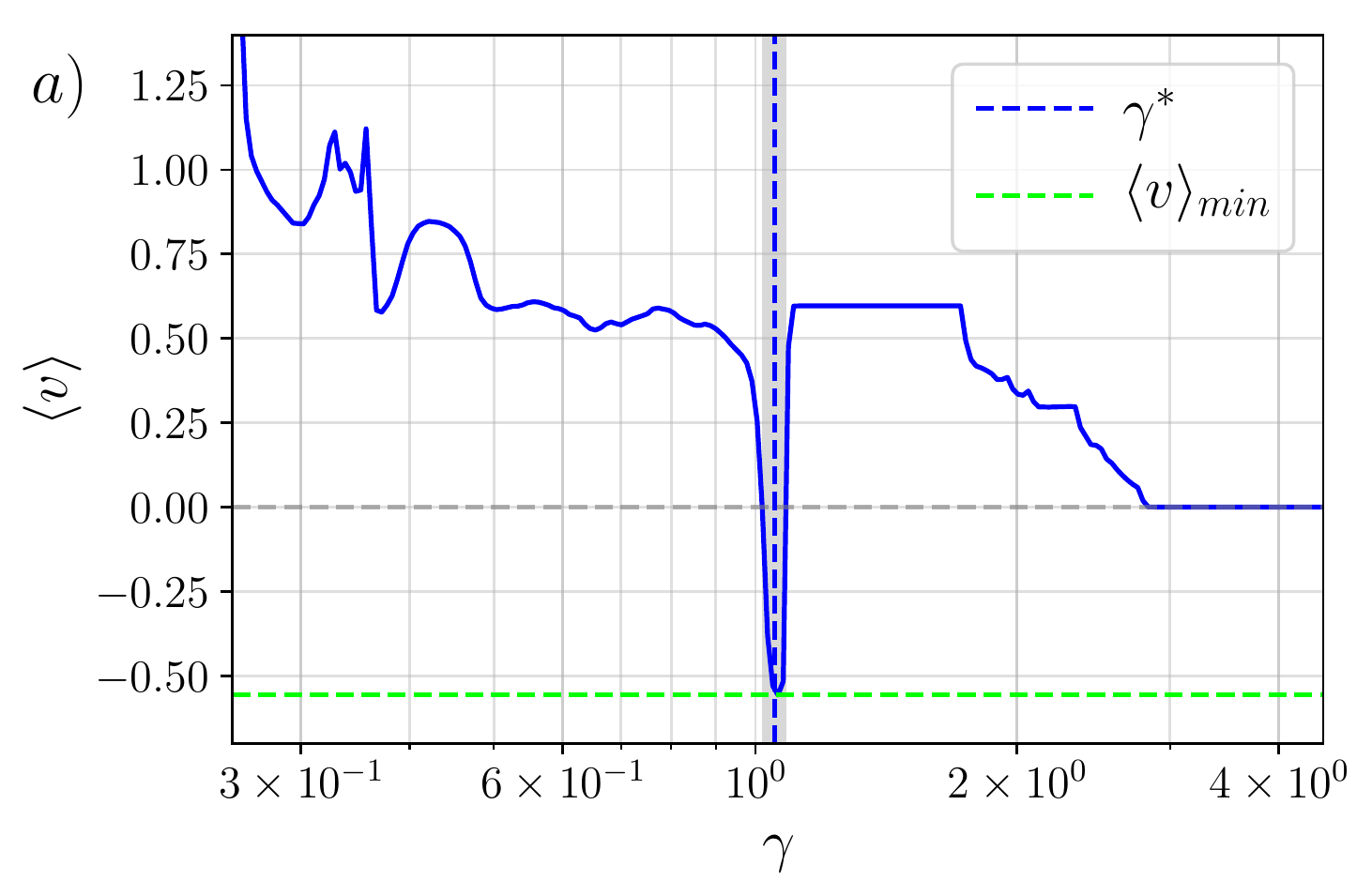}%
	\includegraphics[width=0.49\linewidth]{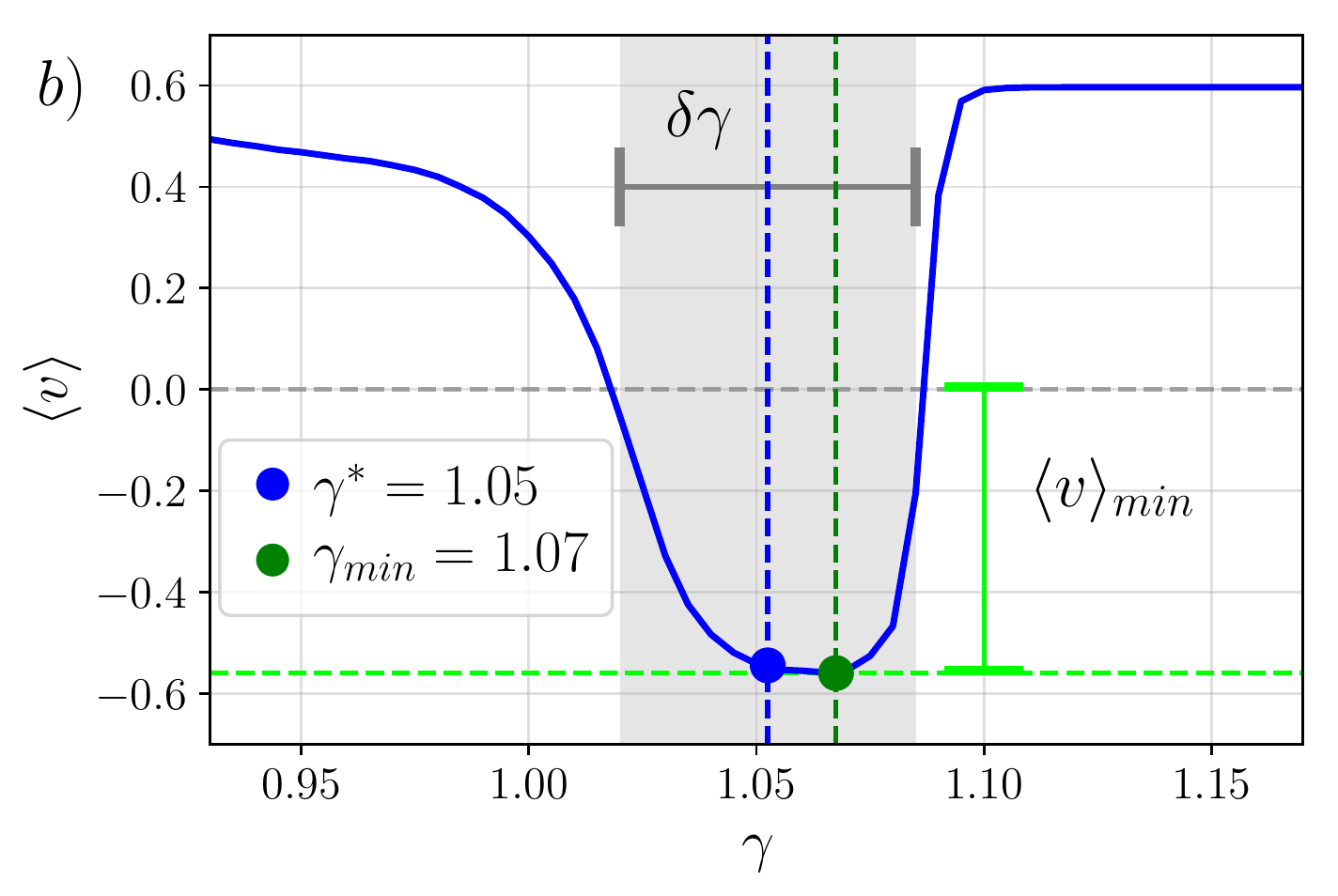}
	\caption{Panel (a): the directed velocity $\langle v \rangle$ versus the friction coefficient $\gamma$. The negative mobility is observed solely for a narrow interval $\delta \gamma$ marked by the grey region. The particle size intended for separation $\gamma^*$ is chosen as the middle of this window. Panel (b) presents a blow up of the interval where the negative mobility emerges. Parameter values: $a = 4.5$, $\omega = 3.75$, $f = 0.7$ and $D = 0.0001$. Note that $a$ and $\omega$ is the same as in Fig. \ref{fig1} and Fig. \ref{fig2}.}
	\label{fig3}
\end{figure*}
\begin{figure*}[t]
	\centering
	\includegraphics[width=0.49\linewidth]{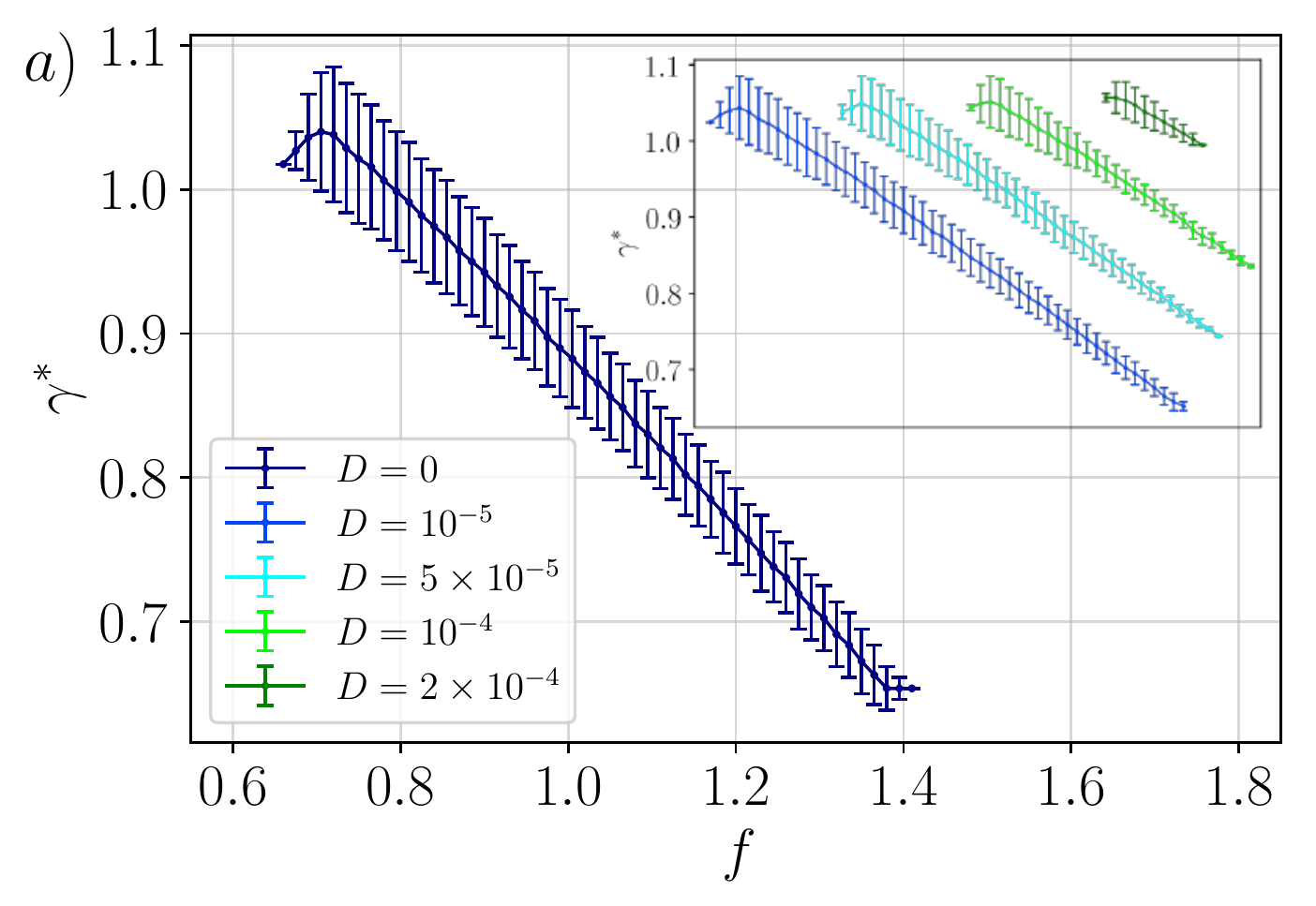}%
	\includegraphics[width=0.49\linewidth]{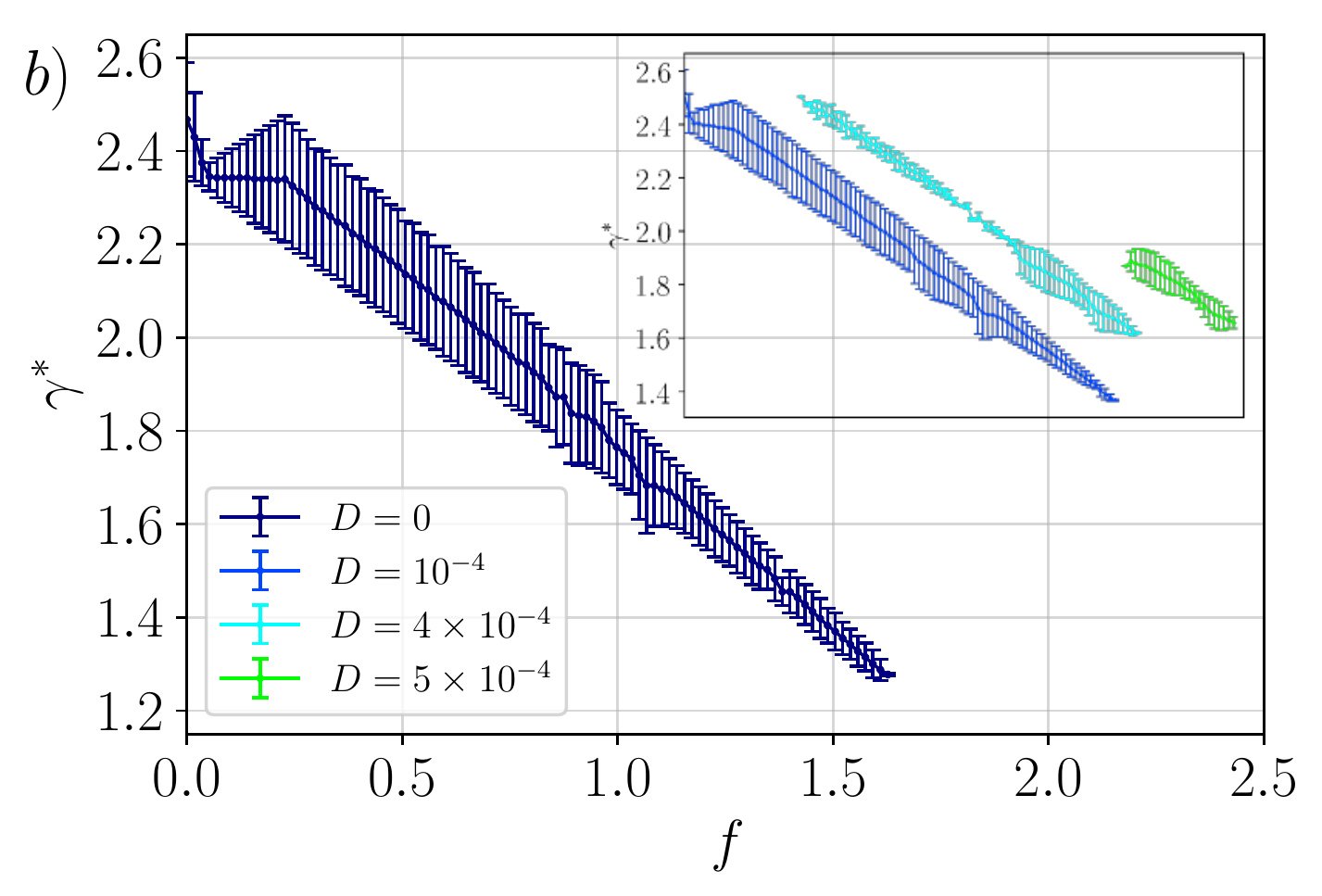}\\
	\includegraphics[width=0.49\linewidth]{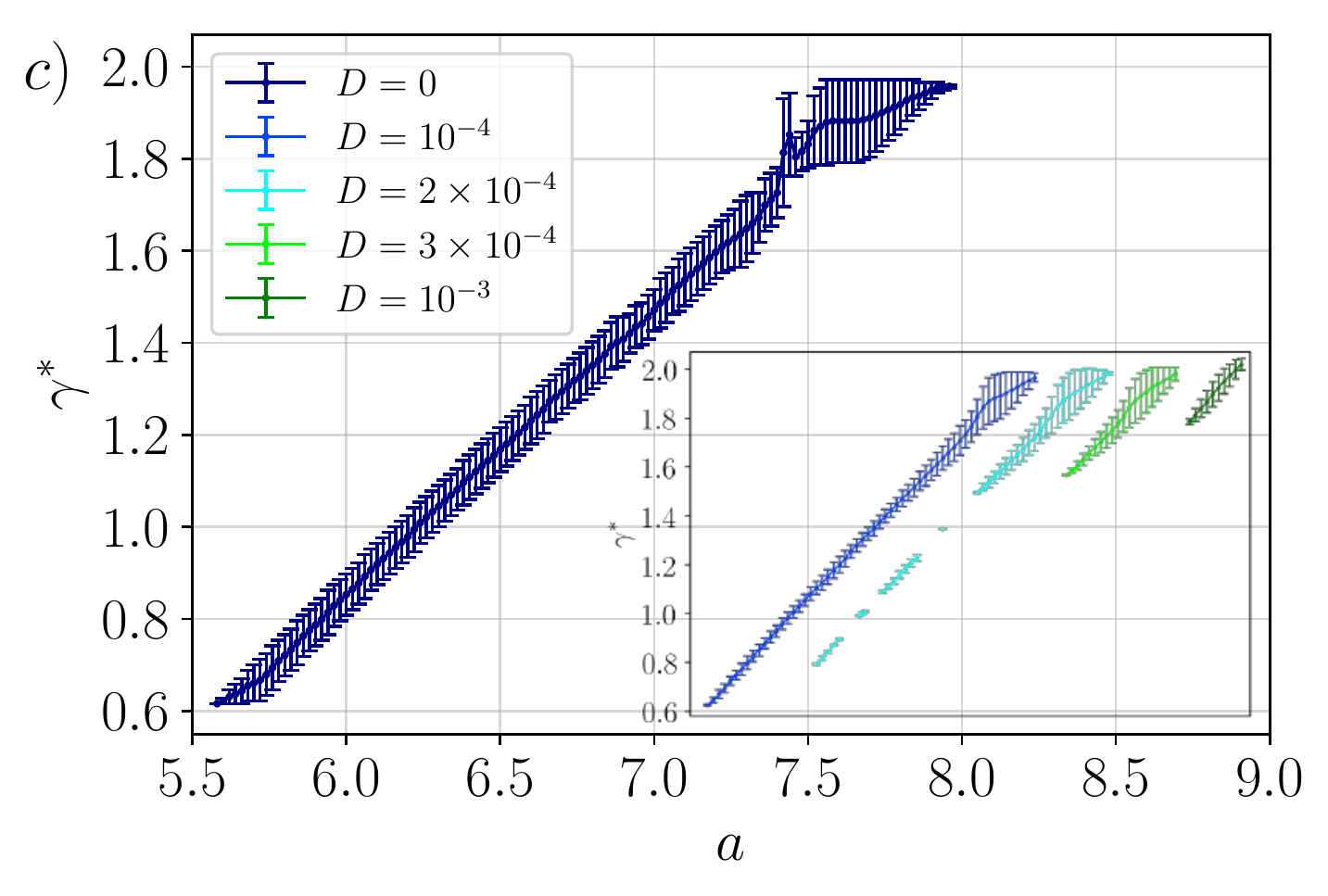}%
	\includegraphics[width=0.49\linewidth]{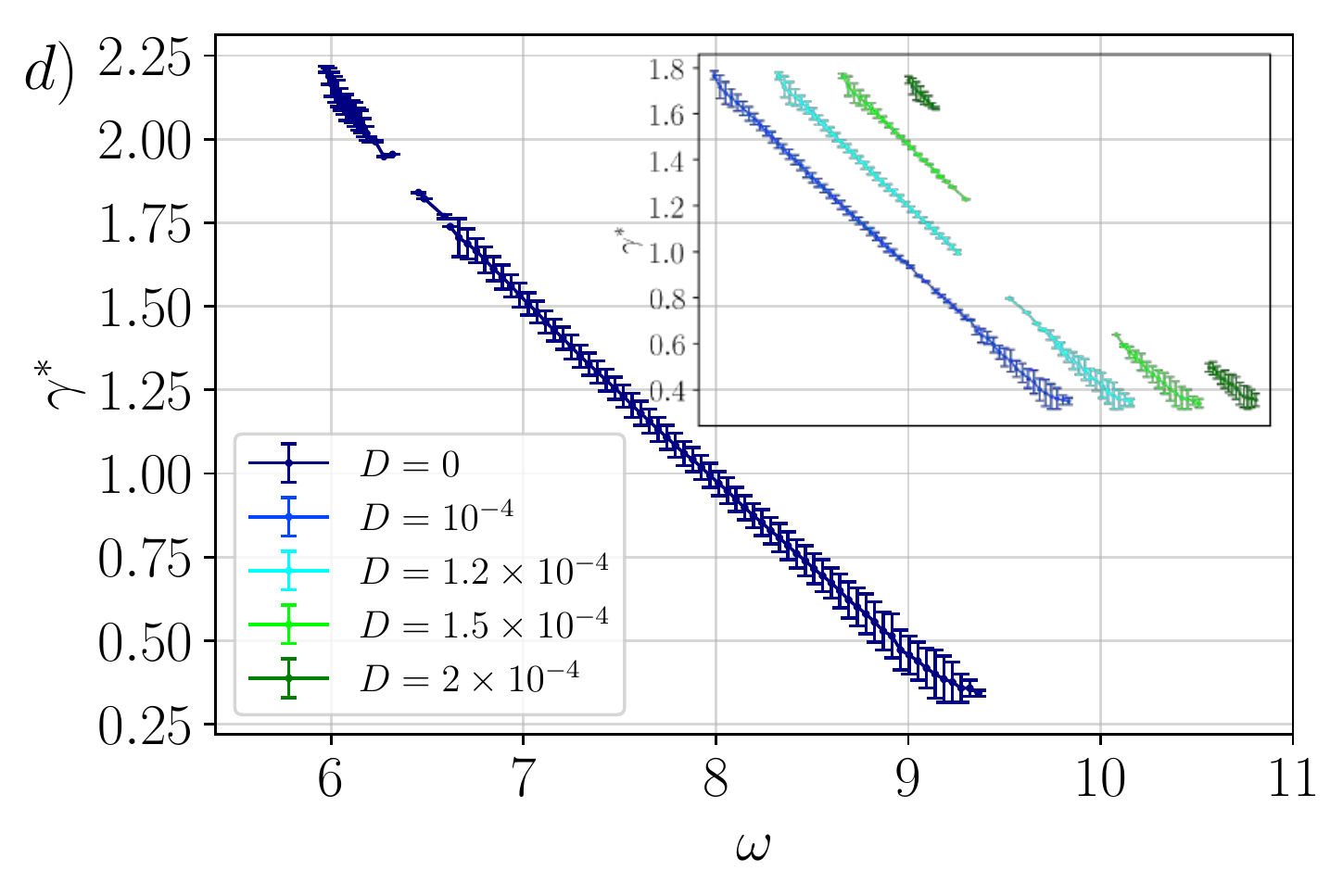}
	\caption{Tunable particle separation. The size targeted for separation $\gamma^*$ is depicted as a function of the parameters of the external force applied to the particle. In panel (a) and (b) versus the static bias $f$, (c) vs the amplitude $a$ and (d) vs the frequency $\omega$. All is depicted for different temperature $D \propto T$ of the system. We stress that all curves shown on the corresponding panels overlap with each other. 
Therefore to visualize the impact of temperature on the separation process in the corresponding insets we schematically show the solutions depicting subsequent $D > 0$ values. CAUTION: the insets do not have labels at $x$-axis to stress the fact that all curves presented there overlap with the deterministic $D = 0$ solution and only $\gamma^*$-axis is relevant for the reader. Other parameters read (a): $a=4.5, \omega=3.75$, (b): $a=11.25, \omega=6.8$, (c): $f=1.5, \omega=4.2$, (d): $f=0.8, a=9.375$.}
	\label{fig5}
\end{figure*}

We exemplify this procedure in Fig. \ref{fig3} where we depict the directed velocity $\langle v \rangle$ versus the friction coefficient $\gamma$ which can be identified with the particle size. As it is illustrated in the panel (a), among many particle sizes corresponding to the friction coefficient $\gamma \in [0.2,4]$ only those with the friction coefficient $\gamma^* \approx 1.05$  will move in the direction opposite $\langle v \rangle < 0$ to the applied bias $f > 0$. Other particles will travel concurrently towards it. As a result, only the particles with $\gamma^* \approx 1.05$ will be extracted from the heterogeneous mixture. Panel (b) presents a magnification of the interval $\delta \gamma$ where the effect of negative mobility emerges. It can be interpreted as a resolution capacity of this method. In this case it reads $\delta \gamma \approx 0.0678$. The selectivity of the proposed separation scheme is impressive as $\delta \gamma/\gamma^* \approx 0.06$ . Moreover, as it is illustrated, typically the velocity $\langle v \rangle$ is noticeably peaked in the interval $\delta \gamma$ where the negative mobility occurs. The mechanism of the latter anomalous transport effect is rooted in the deterministic dynamics and as such thermal fluctuations generally have destructive impact on it \cite{slapik2018cnsns,slapik2019prappl}. Therefore the above outlined particle separation strategy is viable in low to moderate temperature regimes in which the size targeted for isolation $\gamma^*$, i.e. the middle of the interval $\delta \gamma$, coincides well with the the friction coefficient $\gamma_{min}$ for which the directed velocity attains its minimal value $\langle v \rangle_{min} \equiv \langle v \rangle(\gamma_{min})$. It means that not only the selectivity of this method is impressive but also the particle separation process is quick.

Such approach allowed us to distill parameter regimes which reveal a specific functional dependence between the particle size tailored for separation $\gamma^*$ and the parameters of the external force applied to the system thus allowing for tunable particle isolation. In Fig. \ref{fig5} we present the friction coefficient $\gamma^*$ versus the static bias $f$, the amplitude $a$ and the frequency $\omega$ all depicted for different temperature of the system $D \propto T$. The data were obtained from the characteristics $\langle v \rangle(\gamma)$ computed for many values of the control parameter, c.f. Fig. \ref{fig3}. Each dot represents the friction coefficient $\gamma^*$ undergoing the separation process at the fixed $f$, $a$ or $\omega$, see panels (a)-(d). The bars indicate the friction coefficient interval $\delta \gamma$ where the negative mobility emerges. Here we note that the curves on the corresponding panels overlap with the one representing the deterministic solution $D = 0$. Therefore to visualize the impact of temperature on the separation process  in the corresponding insets we schematically show the solutions depicting subsequent $D$ values. The reader may observe that using those tailored  parameters read off from Fig. \ref{fig5} one is able to tune the negative mobility to the particle of a given size $\gamma^*$ by changing solely the static bias $f$ or the amplitude $a$ or the frequency $\omega$. In this way it will be separated from the others possessing positive mobility and thus moving concurrently towards the applied bias $f$.

From the experimental point of view the most convenient way to manipulate the particle separation is presumably by altering the static bias $f$. It is because in many realistic setups it is implemented via the constant external field, e.g. in the microfludic experiments in the form of a spatially uniform electric field which induces the particle electrophoresis \cite{sonker2019}. In most cases its intensity can be changed relatively easily, as opposed to the frequency $\omega$ of the external harmonic driving which often requires complete rebuilding of the experimental setup. We note that the parameter sets reported in Fig. \ref{fig5} allow for the tunable separation of the particles in the regime of moderate-large friction coefficient which is characteristic for low Reynolds numbers, being typical for (sub)micro sized particles immersed in a solution \cite{happel1965}. For example in the panels (a) and (b) corresponding to the isolation driven by the constant force $f$, the friction coefficient $\gamma^* \in [0.6,2.5]$, whereas in (d) when the separation is controlled by the frequency $\omega$ the size $\gamma^* \in [0.25,1.75]$. The reader can observe that the friction coefficient $\gamma^*$ is a decreasing function of the force $f$ and the frequency $\omega$ (panels (a), (b) and (d)) while for the amplitude $a$ it depicts an increasing dependence (panel (c)). Finally, we note that the dimensionless friction coefficient $\gamma$ in Eq. (\ref{gamma}) is influenced not only by the actual friction $\Gamma$ but also by the parameters $\Delta U$ and $L$ of the potential. Therefore, experimentalists may exploit these characteristics of the periodic substrate to further adapt the particle size targeted for separation.

Last but not least, Fig. \ref{fig5} reveals the impact of temperature $D \propto T$ on the tunability of the separation process. As it was stated before, since the negative mobility effect derives from the deterministic dynamics of the system, thermal fluctuations have a destructive influence on it. When temperature increases the regions of negative mobility allowing for the controllable particle separation progressively shrink and eventually vanish completely. Therefore the reported tunability is possible in the widest range of the friction coefficient $\gamma^*$ for low to moderate temperature regimes $D \propto T$. However, for instance, the panel (c) illustrates an interesting effect of thermal fluctuations. An increase of temperature leads not only to shrinking of the range of the friction coefficient $\gamma^*$ for which the negative mobility is observed but also to significant decrease of the intervals where this phenomenon occurs, thereby optimizing the width $\delta \gamma$. It means that then the tunability of the method is limited but selectivity of the separation process increases. We remark that the particle isolation upon harvesting the negative mobility phenomenon is also present for different parameter regimes, however, the range of its tunability proves somewhat smaller.

\begin{figure*}[t]
	\centering
	\includegraphics[width=0.88\linewidth]{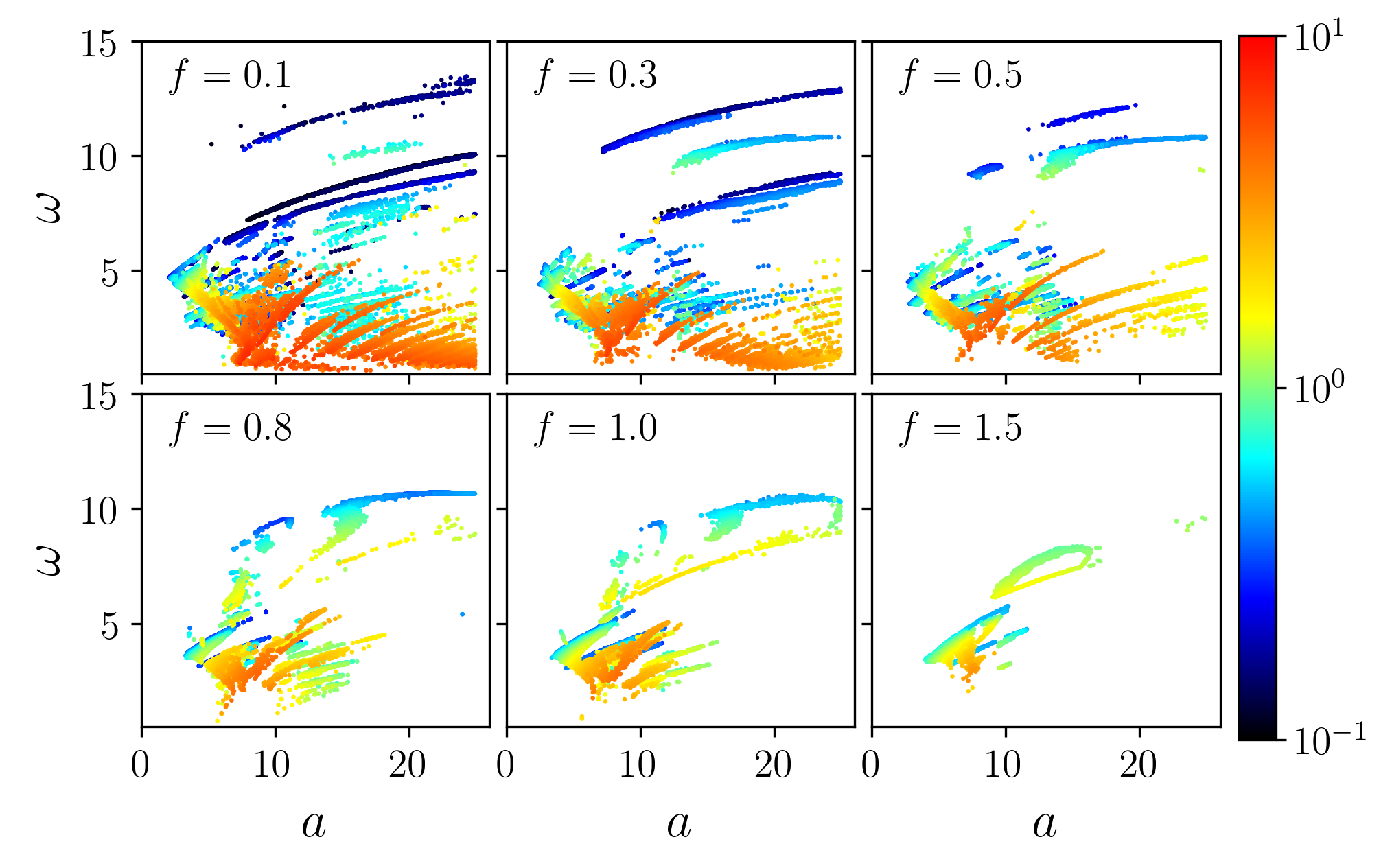}
	\caption{The particle size $\gamma^*$ (color coded scale) targeted for separation via the negative mobility effect as a function of the amplitude $a$ and the frequency $\omega$ for different values of the bias $f$. Thermal noise intensity $D$ is set to zero $D=0$.}
	\label{fig6}
\end{figure*}
\section{Tailoring particle separation}
Now we consider another, complementary issue. Let us assume that we deal with particles of a given size $\gamma^*$ which we want to extract from the heterogeneous mixture. We address the following question: for how many different sizes $\gamma^*$ taken from the extended interval $\gamma^* \in [0.1,10]$ it is possible to find a parameter set $\{a, \omega, f, D\}$ for which the negative mobility effect emerges in the small interval $[\gamma^* - \delta \gamma/2, \gamma^* + \delta \gamma/2]$ around the targeted value $\gamma^*$, therefore allowing its separation in a unique manner, c.f. Fig. \ref{fig3}? We found that in most cases the magnitude $\delta\gamma/\gamma^*$ is equal to a few percent but frequently is even smaller. We note that for a single $\gamma^*$ there might be several parameter regimes fulfilling this condition thus facilitating the choice in realizing the best parameter set. We present the answer of this experimentally and practically relevant question in Fig. \ref{fig6}. The distribution of the size $\gamma^*$ targeted for separation is shown there in the parameter plane of the amplitude $a$ and the frequency $\omega$ for different values of the static bias $f$. The color coded scale displays the friction coefficient $\gamma^*$ value. We observe that small particles can be isolated when the static bias $f$ is likewise small and moderate to large values of the frequency $\omega$. On the other hand, medium and large particles are separated for small frequencies $\omega$. We note that the distribution of the particle size targeted for isolation $\gamma^*$ undergoes a stretching when the static bias $f$ decreases. Moreover, in such a case the range of the particles which might be segregated by harvesting the negative mobility effect is extended as well. Finally, even though the considered panel depicts the deterministic $D = 0$ dynamics, we found that the distribution of $\gamma^*$ depicted there is quite robust with temperature change and survives up to moderate thermal noise intensity, see Fig. \ref{fig1} (b).
\section{Conclusion}
In this work we provide an efficient method for tunable separation of (sub)micro sized particles via the negative mobility phenomenon. The approach presented here requires only a spatially periodic nonlinear structure in combination with an unbiased external time-periodic driving. In this scheme the particle size intended for isolation can be effectively controlled by changing solely the parameters of the external force applied to the system, namely, the static bias or the amplitude or the frequency of the external harmonic driving. The approach can be further adapted to the needs by proper fabrication of the nonlinear potential landscape determined by its barrier height and period. It allows the possibility to not deflect the separated particles along the different angles but to steer them in the opposite direction making the isolation process robust.

Our theoretical predictions should be used as a guide towards physical reality indicating the direction for future theoretical and experimental research. In particular, one needs to carefully consider higher dimensional systems as well as geometrical constraints together with hydrodynamic interaction which in real experiments may play essential role. We expect that such research would potentially lead to implementation of the proposed scheme in a lab-on-chip device, as it has been recently demonstrated for a similar system \cite{luo2016}. We envision that current lithographic techniques with advantageous fabrication costs may be used to develop high throughput separation applications concerning in particular biophysical and biochemical problems. Taking into account recent progress in 3D printing technologies allowing its scaling down to the nanometer range the proposed scheme may have even significant commercial potential in future \cite{chiu2017}.

\section*{Methods}
We employed a weak 2nd order predictor-corrector method \cite{platen} to simulate stochastic dynamics given by Eq. (\ref{dimlessmodel}). We integrated it with the time step scaled by the fundamental period $\mathsf{T} = 2\pi/\omega$ of the external harmonic driving, namely $h = 10^{-2} \times \mathsf{T}$, with the exception of smallest $\omega < 1$ values for which the step was chosen to be $h = 10^{-3} \times \mathsf{T}$. The initial positions $x(0)$ and velocities $v(0)$ were uniformly distributed over the intervals $[0,1]$ and $[-2,2]$, respectively. The directed velocity $\langle v \rangle$ was averaged over the ensemble of $2^{10} = 1024$ trajectories, each starting with a different initial condition according to the above distributions. The number of realisations of stochastic dynamics is not accidental and was chosen carefully to maximise the performance of the numerical simulation, see Ref. \cite{spiechowicz2015cpc} for more details. The time span of the simulations was set to $[0,10^4 \mathsf{T}]$ to guarantee that the directed velocity $\langle v \rangle$ relaxed to its asymptotic long time stationary value.

\section*{Acknowledgment}
This work has been supported by the Grant NCN No. 2017/26/D/ST2/00543 (J. S.).

\section*{Author contributions}
A. S. carried out numerical calculations. All authors contributed to the discussion and analysis of the results. J. S. wrote the manuscript.


\section*{References}

\begin{thebibliography}{99}
\bibitem{yager2006} Yager P., Edwards T., Fu E., Helton K., Nelson K. Microfluidic diagnostic technologies for global public health. \textit{Nature} \textbf{442}, 412 (2004)
\bibitem{korecka2007} Korecka J. A., Verhaagen J., Hol E. M. Cell-replacement and gene-therapy strategies for Parkinson's and Alzheimer's disease. \textit{Regen. Med.} \textbf{2}, 425 (2007)
\bibitem{heffner1978} Heffner R. R., Barron S. A. The early effects of ischemia upon skeletal muscle mitochondria. \textit{J. Neurol. Sci.} \textbf{38}, 295 (1978)
\bibitem{suresh2007} Suresh S. Biomechanics and biophysics of cancer cells. \textit{Acta Mater.} \textbf{55}, 3989 (2007)
\bibitem{bhagat2010} Bhagat A. A., Bow H., Hou S. W., Tan S. J., Han J., Lim C. T. Microfluidics for cell separation. \textit{Med. Biol. Eng. Comput.} \textbf{48}, 999 (2010)
\bibitem{xuan2014} Xuan J., Lee M. L. Size separation of biomolecules and bioparticles using micro/nanofabricated structures. \textit{Anal. Methods} \textbf{6}, 27 (2014)
\bibitem{fehr2015} Fehr A. R. and Perlman S. Coronaviruses: An Overview of Their Replication and Pathogenesis in \textit{Coronaviruses. Methods in Molecular Biology vol 1282} (Humana Press, New York, 2015)
\bibitem{kandel2000} Kandel E. R., Schwartz J. H. and Jessell T. M. \textit{Principles of Neural Science} (McGraw-Hill, 4th ed., New York, 2000)
\bibitem{sajeesh2014} Sajeesh P., Sen A. K. Particle separation and sorting in microfluidic devices: a review. \textit{Microfluid. Nanofluid.} \textbf{17}, 1 (2014)\bibitem{sonker2019} Sonker M., Kim D., Egatz-Gomez A., Ros A. Separation Phenomena in Tailored Micro- and Nanofluidic Environments. \textit{Annu. Rev. Anal. Chem.} \textbf{12}, 475 (2019)
\bibitem{bogunovic2012} Bogunovic L., Eichhorn R., Regtmeier J., Anselmetti D. and Reimann P. Particle sorting by a structured microfluidic ratchet device with tunable selectivity: theory and experiment. \textit{Soft Matter} \textbf{8}, 3900 (2012)
\bibitem{kim2018} Kim D., Luo J., Arriaga E. A. and Ros A. Deterministic Ratchet for Sub-micrometer (Bio)particle Separation. \textit{Anal. Chem.} \textbf{90}, 4370 (2018)
\bibitem{zhang2018} Zhang J. et al. Tunable particle separation in a hybrid dielectrophoresis (DEP)- inertial microfluidic device. \textit{Sensors and Actuators B} \textbf{267}, 14 (2018)
\bibitem{eichhorn2002} Eichhorn R., Reimann P., H\"anggi P. Brownian motion exhibiting absolute negative mobility. \textit{Phys. Rev. Lett.} \textbf{88}, 190601 (2002)
\bibitem{machura2007} Machura {\L}., Kostur M., Talkner P., {\L}uczka J., H\"anggi P. Absolute negative mobility induced by thermal equilibrium fluctuations. \textit{Phys. Rev. Lett.} \textbf{98}, 040601 (2007)
\bibitem{spiechowicz2014pre} Spiechowicz J., H\"anggi P., {\L}uczka J. Brownian motors in the microscale domain: Enhancement of efficiency by noise. \textit{Phys. Rev. E} \textbf{90}, 032104 (2014)
\bibitem{spiechowicz2019njp} Spiechowicz J., H\"anggi P. and {\L}uczka J. Coexistence of absolute negative mobility and anomalous diffusion, \textit{New J. Phys.} \textbf{21}, 083029 (2019)
\bibitem{ros2005} Ros A., Eichhorn R., Regtmeier J., Duong T. T., Reimann P., Anselmetti D. Absolute negative mobility. \textit{Nature} \textbf{436}, 928 (2005)
\bibitem{eichhorn2010} Eichhorn R., Regtmeier J., Anselmetti D., Reimann P. Negative mobility and sorting of colloidal particles. \textit{Soft Matter} \textbf{6}, 1858 (2010)
\bibitem{luo2016} Luo J., Muratore K., Arriaga E., Ros A. Deterministic Absolute Negative Mobility for Micro- and Submicrometer Particles Induced in a Microfluidic Device. \textit{Anal. Chem.} \textbf{88}, 5920 (2016)
\bibitem{slapik2019prl} Slapik A., {\L}uczka J., H\"anggi P., Spiechowicz J. Tunable Mass Separation via Negative Mobility. \textit{Phys. Rev. Lett.} \textbf{122}, 070602 (2019)
\bibitem{hanggi2009} H\"anggi P., Marchesoni F. Artificial Brownian motors: Controlling transport on the nanoscale. \textit{Rev. Mod. Phys.} \textbf{81}, 387 (2009)
\bibitem{marconi2008fluctuation} Marconi U. M. B., Puglisi A., Rondoni L. and Vulpiani A. Fluctuation-dissipation: response theory in statistical physics, \textit{Phys. Rep.} \textbf{461}, 111 (2008)
\bibitem{spiechowicz2017scirep} Spiechowicz J., {\L}uczka J. Subdiffusion via dynamical localization induced by thermal equilibrium fluctuations. \textit{Sci. Rep.} \textbf{7}, 16451 (2017)
\bibitem{spiechowicz2019chaos} Spiechowicz J. and {\L}uczka J. SQUID ratchet: Statistics of transitions in dynamical localization, \textit{Chaos} \textbf{29}, 013105 (2019)
\bibitem{reimann2001} Reimann P., Van den Broeck C., Linke H., H\"anggi P., Rubi J. M., Perez-Madrid A. Giant acceleration of free diffusion by use of tilted periodic potentials. \textit{Phys. Rev. Lett.} \textbf{87}, 010602 (2001)
\bibitem{spiechowicz2015chaos} Spiechowicz J., {\L}uczka J. Josephson phase diffusion in the superconducting quantum interference device ratchet. \textit{Chaos} \textbf{25}, 053110 (2015)
\bibitem{spiechowicz2020pre} Spiechowicz J. and {\L}uczka J. Diffusion in a biased washboard potential revisited. \textit{Phys. Rev. E} \textbf{101}, 032123 (2020)
\bibitem{landau_hydro} Landau L. D. and Lifshitz E. M. \textit{Fluid Mechanics} (Butterworth-Heinemann, 2nd ed., 1987)
\bibitem{spiechowicz2016scirep} Spiechowicz J., {\L}uczka J., H\"anggi P. Transient anomalous diffusion in periodic systems: ergodicity, symmetry breaking and velocity relaxation. \textit{Sci. Rep.} \textbf{6}, 30948 (2016)
\bibitem{spiechowicz2015cpc} Spiechowicz J., Kostur M., Machura {\L}. GPU accelerated Monte Carlo simulation of Brownian motors dynamics with CUDA. \textit{Comp. Phys. Commun.} \textbf{191}, 140 (2015)
\bibitem{speer2007} Speer D., Eichhorn R., Reimann P. Transient chaos induces anomalous transport properties of an underdamped Brownian particle. \textit{Phys. Rev. E} \textbf{76}, 051110 (2007)
\bibitem{nagel2008} Nagel J., Speer D., Gaber T., Sterck A., Eichhorn R., Reimann P., Ilin K., Siegel M., Koelle D., Kleiner R. Observation of Negative Absolute Resistance in a Josephson Junction. \textit{Phys. Rev. Lett.} \textbf{100}, 217001 (2008)
\bibitem{landau_stat} Landau L. D. and Lifshitz E. M. \textit{Statistical Physics, Part 1} (Butterworth-Heinemann, 3rd ed., 1980)
\bibitem{slapik2018cnsns} Slapik A., {\L}uczka J. and Spiechowicz J. Negative mobility of a Brownian particle: Strong damping regime. \textit{Commun. Nonlinear Sci. Numer. Simul.} \textbf{55}, 316 (2018)
\bibitem{slapik2019prappl} Slapik A., {\L}uczka J. and Spiechowicz J. Temperature-induced tunable particle separation, \textit{Phys. Rev. Appl.} \textbf{12}, 054002 (2019)
\bibitem{kostur2008} Kostur M., Machura {\L}., Talkner P., H\"anggi P. and {\L}uczka J. Anomalous transport in biased ac-driven Josephson junctions: Negative conductances, \textit{Phys. Rev. B} \textbf{77}, 104509 (2008)
\bibitem{happel1965} Happel J. and Brenner H. \textit{Low Reynolds Number Hydrodynamics} (Prentice Hall, Englewood Cliffs, NJ, 1965)
\bibitem{chiu2017} Chiu D. T., Carlo D. Di, Doyle P. S., Hansen C., Maceiczyk R. M., Wootton R. C. Small but perfectly formed? Successes, challenges and opportunities for microfluidics in the chemical and biological sciences. \textit{Chemistry} \textbf{2}, 201 (2017)
\bibitem{platen} Platen E. and Bruti-Liberati N. \textit{Numerical Solution of Stochastic Differential equations with Jumps in Finance} in Stochastic Modelling
and Applied Probability (Springer, Berlin, 2010)

\end{thebibliography}
\end{document}